\documentclass[preprint2]{aa}
\usepackage{lscape}
\usepackage{graphicx}
\usepackage{longtable}
\usepackage{float}

\usepackage{rotating}
\usepackage{graphicx}
\usepackage{booktabs}
\begin{document}


\title {
H$_2$CS deuteration maps towards the pre-stellar core L1544\thanks{Based on observations carried out with the IRAM 30m telescope. IRAM is supported by INSU/CNRS (France), MPG (Germany) and IGN (Spain)}
}


\author{S. Spezzano\inst{1}  \and O. Sipil\"a\inst{1}  \and P. Caselli\inst{1}  \and S. S. Jensen\inst{1} \and S. Czakli\inst{1}  \and L. Bizzocchi\inst{2} \and J. Chantzos\inst{1} \and G. Esplugues\inst{3} \and A. Fuente\inst{3} \and F. Eisenhauer\inst{1} }  

\institute{Max-Planck-Institut f\"ur Extraterrestrische Physik, Giessenbachstrasse 1, 85748 Garching, Germany \and University of Bologna, Dipartimento di Chimica “Giacomo Ciamician”, Bologna, Italy \and Observatorio Astron\'omico Nacional (OAN), Alfonso XII, 3, 28014, Madrid}

\abstract 
{Deuteration is a crucial tool to understand the complexity of interstellar chemical processes, especially when they involve the interplay of gas-phase and grain-surface chemistry. In the case of multiple deuteration, comparing observation with the results of chemical modelling is particularly effective to study how molecules are inherited in the different stages within the process of star and planet formation.}{We aim to study the the D/H ratio in H$_2$CS across the prototypical pre-stellar core L1544. This study allows us to test current gas–dust chemical models involving sulfur in dense cores.}{We present here single-dish observations of H$_2$CS, HDCS and D$_2$CS with the IRAM 30m telescope. We analyse their column densities and distributions, and compare these observations with gas–grain chemical models. The deuteration maps of H$_2$CS in L1544 are compared with the deuteration maps of methanol, H$_2$CO, N$_2$H$^+$ and HCO$^+$ towards the same source. Furthermore, the single and double deuteration of H$_2$CS towards the dust peak of L1544 is compared with H$_2$CO and $c$-C$_3$H$_2$. The difference between the deuteration of these molecules in L1544 is discussed and compared with the prediction of chemical models.}{The maximum deuterium fractionation for the first deuteration of H$_2$CS is N(HDCS)/N(H$_2$CS)$\sim$30$\%$ and is located towards the north-east at a distance of about 10000 AU from the dust peak. While for $c$-C$_3$H$_2$ the first and second deuteration have a similar efficiency, for H$_2$CS and H$_2$CO the second deuteration is more efficient, leading to D$_2$CX/HDCX$\sim$100$\%$ (with X= O or S).}
{Our results imply that the large deuteration of H$_2$CO and H$_2$CS observed in protostellar cores as well as in
comets is likely inherited from the pre-stellar phase. However, the comparison with state-of-the-art chemical models suggests that the reaction network for the formation of the doubly deuterated H$_2$CS and H$_2$CO it is not be complete yet.}{}
%{} {aa}{} {.}

\titlerunning{H$_2$CS deuteration maps}
\maketitle

\section{Introduction}
A considerable enhancement in molecular deuteration is observed towards the early stages of star formation.  Despite the cosmic deuterium abundance relative to hydrogen being $\sim$1.5$\times$10$^{-5}$ \citep{linsky03}, the abundances of deuterated molecules relative to the correspondent main species range from $\sim$20\% for N$_2$H$^+$ towards pre-stellar cores to 10\% for methanol and formaldehyde in protostellar cores \citep{parise06}.
These enhancements are a consequence of the exothermic reaction

\begin{equation}
   \rm H_3^+ + HD \rightleftarrows H_2D^+ + H_2 + 230 K,
\end{equation}

and successive deuterations up to the formation of D$_3^+$ (see \citealt{ceccarelli14} and references therein). The deuterated isotopologues of H$_3^+$ can also dissociatively recombine with electrons, enhancing the atomic D/H ratio in the gas phase \citep{roberts03}. While H$_2$D$^+$, and the multiply deuterated isotopologues of H$_3^+$, are deuterating molecules in the gas phase, the enhanced atomic D/H ratio is transferred to grains where it deuterates the molecules on the surface. 
Isotopic fractionation, and in particular the deuterium fractionation, is a powerful tool to study the evolution of material during the process of star and planetary system formation. Observations of ortho and para H$_2$D$^+$, for example, have allowed to derive the age of a core forming a Sun-like star \citep{bruenken14}. Furthermore, observing and modelling water and its deuterated isotopologues has established that a substantial fraction of the water in the Solar System is inherited from the pre-stellar core where the Sun formed \citep{cleeves14, vandishoeck21}. The case of water has highlighted how powerful it is to use deuteration as a probe of inheritance, and it is important to note that the conclusive evidence was found in the high abundances of the doubly deuterated water. Molecules with the possibility of multiple deuteration provide in fact crucial constraints not only to the deuteration processes involved, but also to the formation of the main species. This is essential particularly in the case of molecules like H$_2$CO and H$_2$CS that are formed with an interplay of of gas-phase and grain-surface chemistry.

H$_2$CS, likewise H$_2$CO, is formed both in the gas phase (mainly from atomic S) and on the surface of dust grains by addition of hydrogen atoms on CS. H$_2$CS has been observed towards cold molecular clouds, protostellar cores, hot cores, circumstellar envelopes and protoplanetary disks \citep{sinclair73, vastel18, agundez08,legal19, drozdovskaya18}. Its deuterated isotopologues have also been observed towards starless and prostostellar cores \citep{marcelino05, drozdovskaya18}.

The low-mass pre-stellar core L1544, located in the Taurus molecular cloud at 170 pc \citep{galli19}, is a well-studied object. It is centrally concentrated \citep{wt99} and shows signs of contraction motion. Its central density is between $\sim$10$^6$ and $\sim$10$^7$ cm$^{-3}$ and the central temperature is $\sim$6 K \citep{Keto10a,crapsi07}. The core exhibits a high degree of CO freeze-out, and a high level of deuteration towards its center \citep{crapsi05}. It is chemically rich, showing spatial inhomogeneities in the distribution of molecular emission \citep{spezzano17, redaelli19, chacon19}.
The sulfur chemistry towards the dust peak of the pre-stellar core L1544 has been presented in \cite{vastel18}, where it is shown that the sulfur-bearing species in L1544 are emitting from an external layer ($\sim$ 10$^4$ AU from the core centre). Furthermore, only a fraction of a percent of the cosmic sulfur fractional abundance w.r.t. total H nuclei (1.5$\times$10$^{-5}$) is needed to reproduce the observations, confirming that sulfur is highly depleted in the dense interstellar medium \citep{laas19}.

In this paper we present the first deuteration maps of H$_2$CS towards the pre-stellar core L1544. In Section 2 we present the observations. The analysis is described in Section 3, and the chemical modelling is presented in Section 4. We discuss the results in Section 5 and summarise our conclusions in Section 6.

\begin{table*}{}
\caption{Spectroscopic parameters of the observed lines}
\label{table:parameters}
%\scalebox{1}{
\begin{tabular}{ccccc}
\hline\hline \\[-2ex]
Molecule & Transition & Rest frequency & $E_\text{up}$ & $n_\text{crit}$ (at 10\,K)   \\
&    $J_{K_a,K_c}$   &(MHz)   & (K) &(cm$^{-3}$)                                      \\[0.5ex]
\hline \\[-2ex]
H$_2$CS   &  $3_{0,3}-2_{0,2}$ &   103040.447(1)       &  9.9    & $1.5\times 10^5$ \\
HDCS      &  $3_{0,3}-2_{0,2}$ &    92981.60(2)        &  8.9    &     --           \\
D$_2$CS   &  $3_{0,3}-2_{0,2}$ &    85153.92(5)        &  8.1    &     --           \\
\hline
\end{tabular}
\tablefoot{Numbers in parentheses denote $1\sigma$ uncertainties in unit of the last quoted digit. n$_{crit}$ is the critical density of the transition.}
%\textit{Notes:}The have not been computed yet. However, we are observing the same transition for all isotopologues, and we can assume that the critical densities for the 3$_{0,3}$-2$_{0,2}$ transition of HDCS and D$_2$CS are comparable with the one of H$_2$CS.
\end{table*}

\section{Observations}
\begin{figure*}
\begin{center}
\includegraphics[width=19cm]{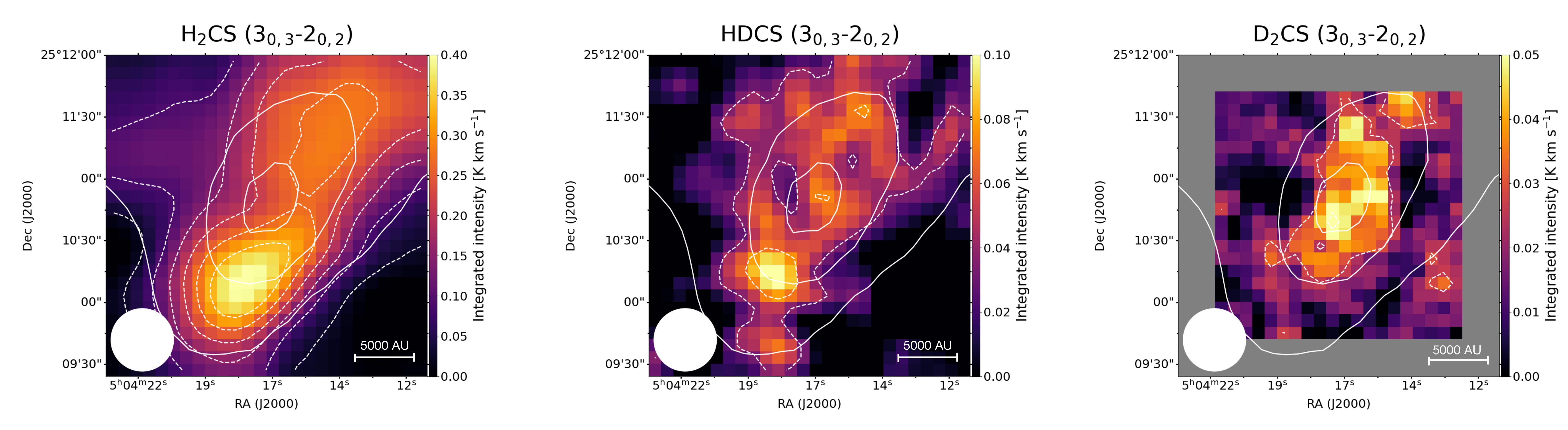}% This is a *.eps file
\end{center}
\caption{Integrated intensity maps of the 3$_{0,3}$-2$_{0,2}$ transitions of H$_2$CS, HDCS, and D$_2$CS towards the inner 2.5$^\prime$ $\times$2.5$^\prime$ of L1544. All maps have been smoothed to 30.5$^{\prime\prime}$, and the beam is shown at the bottom left of each map. The solid white contours are the 30\%, 60\% and 90\% of the peak intensity of the N(H$_2$) map of L1544 computed from {\em Herschel}/SPIRE data \citep{spezzano16}. The dashed white contours indicate the 3$\sigma$ integrated emission with steps of 3$\sigma$ ({\normalfont rms}$_{H_2CS}$= 10 mK km s$^{-1}$, {\normalfont rms}$_{HDCS}$=12 mK km s$^{-1}$, {\normalfont rms}$_{D_2CS}$= 9 mK km s$^{-1}$).}
\label{fig:integrated_intensity}
\end{figure*}

 The emission maps of H$_2$CS, HDCS, and D$_2$CS towards L1544 were obtained using the IRAM 30m telescope (Pico Veleta, Spain) in 2 different observing runs in 2013 and 2015, and are shown in Figure~\ref{fig:integrated_intensity}. The size of the D$_2$CS map shown in Figure~\ref{fig:integrated_intensity} is smaller with respect to the size of the H$_2$CS and HDCS maps because, given the weakness of the D$_2$CS line, in 2015 a deeper integration for this line was performed towards the inner region of L1544. The spectra of the three isotopologues extracted towards the dust peak of L1544 are shown in Figure~\ref{fig:spectra}. We performed a 2.5$^\prime$ $\times$2.5$^\prime$ on-the-fly (OTF) map centred on the source dust emission peak ($\alpha _{2000}$ = 05$^h$04$^m$17$^s$.21,  $\delta _{2000}$ = +25$^\circ$10$'$42$''$.8). We used position switching with the reference position set at (-180$^{\prime \prime}$,180$^{\prime\prime}$) offset with respect to the map centre. The observed transitions are summarised in Table \ref{table:parameters}. The EMIR E090 receiver was used with the Fourier transform spectrometer backend (FTS) with a spectral resolution of 50 kHz. The mapping was carried out in good weather conditions ($\tau_{225GHz}$ $\sim$ 0.3) and a typical system temperature of T$_{sys}$ $\sim$ 90-150 K. The data processing was done using the GILDAS software \citep{pet05}. The emission maps of H$_2$CS and HDCS have been smoothed to the beam size of D$_2$CS (30.5$^{\prime\prime}$). All maps have been gridded to a pixel size of 6$''$ with the CLASS software in the GILDAS package, which corresponds to 1/5 of the beam size. The integrated intensity maps shown in Figure~\ref{fig:integrated_intensity} have been computed in the 6.9-7.6 km s$^{-1}$ velocity range. To compute the column densities, we used the forward efficiency $F_{eff}$=0.95 and main beam efficiency B$_{eff}$=0.76 to convert the T$_A^*$ temperature scale into the T$_{MB}$ temperature scale. The velocity rest frame used in this work is the local standard of rest (lsr).

\section{Analysis}
\subsection{Deuteration maps}
The column density maps of each H$_2$CS isotopologue (shown in Figure~\ref{fig:H2CS_cd}) have been computed using the formula reported in \cite{Mangum15}, assuming that the source fills the beam, and the excitation temperature T$_{ex}$ is constant:

\begin{equation} 
 N_{tot} = \frac{8\pi\nu^3Q_{rot}(T_{ex})W}{c^3A_{ul}g_u}\frac{e^{\frac{E_u}{kT}}}{J(T_{ex}) -  J(T_{bg})},\\
\end{equation}

\noindent
where $J(T) = {\frac{h\nu}{k}}(e^{\frac{h\nu}{kT}}-1)^{-1}$ is the
source function in Kelvin, $k$ is the Boltzmann constant, $\nu$
is the frequency of the line, $h$ is the Planck constant, $c$ is the speed
of light, $A_{ul}$  is the Einstein
coefficient of the transition, $W$ is the integrated intensity, $g_u$ is the degeneracy of the upper state, $E_u$ is the
energy of the upper state, $Q_{rot}$ is the partition function of the molecule at the given temperature $T_{ex}$. $T_{bg}$ is the
background (2.7 K) temperature. 
Following the results of the MCMC analysis reported in \cite{vastel18}, we assumed $T_{ex}$ = 12.3, 6.8 and 9.3 K for H$_2$CS, HDCS and D$_2$CS, respectively. While computing the column density across the core, the excitation temperature was kept constant for each isotopologue.
The error introduced by using a constant excitation temperature to calculate the column density map across a pre-stellar core has been found negligible in a previous study of L1544 (see in the Appendix of \citealt{redaelli19}). \\

\begin{figure*}
\begin{center}
\includegraphics[width=18cm]{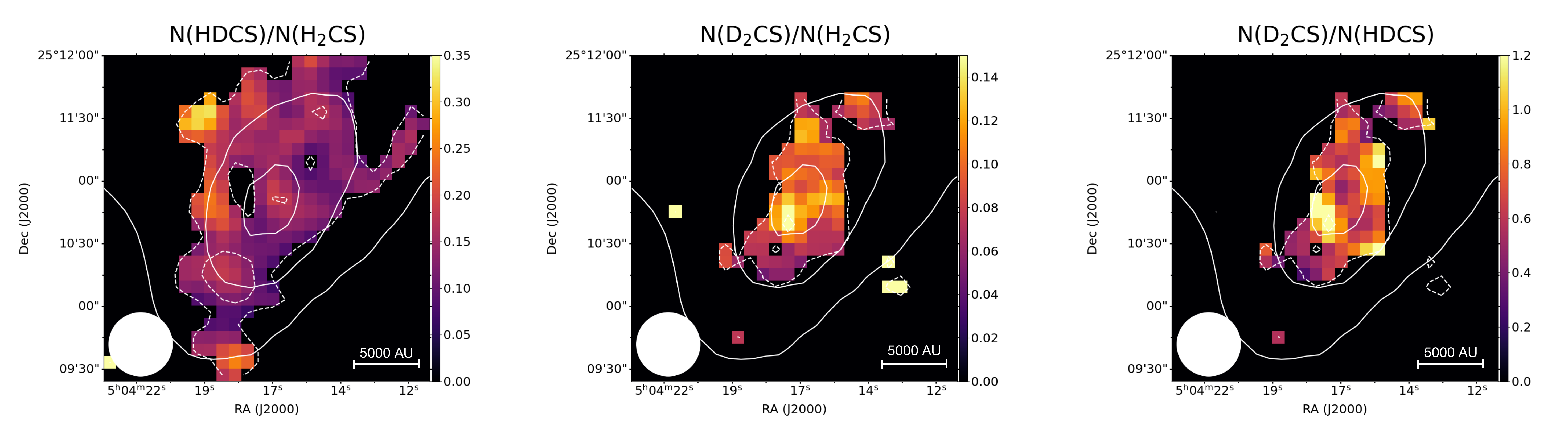}% This is a *.eps file
\end{center}
\caption{Deuteration maps of H$_2$CS towards L1544. The column density ratio has been computed only in pixels where both molecules have been observed at least at a 3$\sigma$ level. The solid white contours are the 30\%, 60\% and 90\% of the peak intensity of the N(H$_2$) map of L1544 computed from {\em Herschel}/SPIRE data. The dotted white contours indicate the 3$\sigma$ integrated emission contour for HDCS in the left panel, and of D$_2$CS in the central and right panel. }
\label{fig:H2CS_ratio}
\end{figure*}

The deuteration maps are shown in Figure~\ref{fig:H2CS_ratio}.
Towards the dust peak N(HDCS)/N(H$_2$CS)$\sim12\pm2\%$, N(D$_2$CS)/N(H$_2$CS)$\sim12\pm2\%$ and N(D$_2$CS)/N(HDCS)$\sim100\pm16\%$, consistent with the values reported in \cite{vastel18}, where the column densities have been derived assuming constant excitation temperature with the MCMC as well as with the rotational diagram method, using three rotational transitions for each of the H$_2$CS isotopologues. The column density ratios involving D$_2$CS have been detected with a signal to noise larger than 3 only towards a 30$''$ by 60$''$ region around the center of L1544, and show an increase of deuteration towards the dust peak. However, given the small coverage, it is difficult to draw conclusions on spatial variations of the efficacy of the second deuteration of H$_2$CS.
The N(HDCS)/N(H$_2$CS) map shows that the deuteration peak for H$_2$CS in L1544 is located towards the north-east at a distance of about 10000 AU ($\sim$60$^{\prime\prime}$) from the dust peak, where N(HDCS)/N(H$_2$CS)$\sim27\pm7\%$. 
To test the effects of using different excitation temperatures for the three isotopologues on our results, we have computed the column density maps and deuteration maps assuming $T_{ex}$ = 9.3 K for all isotopologues, see Figure~\ref{fig:H2CS_cd9.4}. The difference in the corresponding column densities is rather small, with N(H$_2$CS) showing the largest variation ($\sim$20$\%$). However, the deuteration maps do not show significant changes, and most importantly the single deuteration peak is still located in the same position as in the map shown in Figure~\ref{fig:H2CS_ratio}.

\begin{table}
        \centering
        \caption{Initial abundances (with respect to $n_{\rm H} \approx 2\,n({\rm H_2})$) used in the chemical modeling.}
        \begin{tabular}{l|l}
                \hline
                \hline
                Species & Abundance\\
                \hline
                $\rm H_2$ & $5.00\times10^{-1}\,^{(a)}$\\
                $\rm He$ & $9.00\times10^{-2}$\\
                $\rm C^+$ & $1.20\times10^{-4}$\\
                $\rm N$ & $7.60\times10^{-5}$\\
                $\rm O$ & $2.56\times10^{-4}$\\
                $\rm S^+$ & $8.00\times10^{-8}$\\
                $\rm Si^+$ & $8.00\times10^{-9}$\\
                $\rm Na^+$ & $2.00\times10^{-9}$\\
                $\rm Mg^+$ & $7.00\times10^{-9}$\\
                $\rm Fe^+$ & $3.00\times10^{-9}$\\
                $\rm P^+$ & $2.00\times10^{-10}$\\
                $\rm Cl^+$ & $1.00\times10^{-9}$\\
                \hline
        \end{tabular}
        \label{tab:initialabundances}
        \tablefoot{$^{(a)}$ The initial $\rm H_2$ ortho/para ratio is $1 \times 10^{-3}$.}
\end{table}

\begin{table*}
\caption{Column density of the normal, singly, and doubly deuterated isotopologues of $c$-C$_3$H$_2$, H$_2$CO, and H$_2$CS, and their deuteration ratios towards the dust peak of L1544.}
\label{table:ratios}
\scalebox{0.95}{
\begin{tabular}{cc|cc|cc}
\hline\hline \\[-2ex]
\multicolumn{6}{c}{Column densities (10$^{12}$ cm$^{-2}$)}\\
 \hline \\[-2ex]
  $c$-C$_3$H$_2$ &37(1)&H$_2$CO&36(23)&H$_2$CS&6.9(6)\\
  $c$-C$_3$HD &6.2(3)&HDCO&1.30(9)&HDCS&0.8(1)\\
   $c$-C$_3$D$_2$ &0.66(2)&D$_2$CO&1.5(3)&D$_2$CS&0.80(8)\\
 \hline \\[-2ex]
   \multicolumn{6}{c}{Column density ratios}\\
 \hline \\[-2ex]
 $c$-C$_3$HD/ $c$-C$_3$H$_2$&17(1)\%&HDCO/H$_2$CO&4(2)\%&HDCS/H$_2$CS&12(2)\% \\
  $c$-C$_3$D$_2$/ $c$-C$_3$H$_2$&1.7(1)\%&D$_2$CO/H$_2$CO&4(3)\%&D$_2$CS/H$_2$CS&12(2)\%\\
  $c$-C$_3$D$_2$/ $c$-C$_3$HD&10(1)\%&D$_2$CO/HDCO&115(10)\%&D$_2$CS/HDCS&100(16)\%\\
 \hline
\end{tabular}
}
\tablefoot{The column densities of H$_2$CO and $c$-C$_3$H$_2$ have been calculated from the column density of the $^{13}$C isotopologues, assuming a $^{12}$C/$^{13}$C ratio of 68.
Numbers in parentheses denote $1\sigma$ uncertainties in units of the last quoted digit. References: $c$-C$_3$H$_2$ and isotopologues from Spezzano et al. 2013, H$_2$CO and isotopologues from Chac\'on-Tanarro et al. 2019, H$_2$CS and isotopologues from this work.  \\
}

\end{table*}

\section{Comparison with chemical models}
\label{chemical_models}

To investigate whether the observed trends in the deuteration of H$_2$CS and H$_2$CO can be understood in the context of the current knowledge of deuterium chemistry in the ISM, we have run a set of gas-grain chemical simulations attempting to reproduce the column densities and column density ratios observed toward L1544. For this, we used our chemical model which includes an extensive description of deuterium and spin-state chemistry; the main features of the chemical code and the chemical networks are described in detail in \citet{Sipila15a, Sipila15b, Sipila19b}, and are omitted here for brevity. We assume monodisperse spherical grains with a radius of 0.1\,$\mu$m and use the initial abundances displayed in Table~\ref{tab:initialabundances}. The simulation results discussed below correspond to a two-phase chemical model, i.e., one where the ice on the grain surface layer is treated as a single active layer.

For the present work, we have run a set of single-point chemical simulations to check the effect of the volume density on the deuterium fractionations. In these simulations, the temperature and visual extinction are set to ``standard'' values for starless cores: $T_{\rm gas} = T_{\rm dust} = 10\,\rm K$, $A_{\rm V} = 10 \, \rm mag$. We have also run a core simulation using the physical model for L1544 presented by \citet{Keto10a}; the physical model was divided into concentric shells and chemical simulations were run in each shell to produce time-dependent radially varying abundance profiles. Essentially the same modeling procedure was recently used in \citet{redaelli21} to investigate the $r$-dependence of the cosmic-ray ionization rate in L1544. It was found in that paper that observations of the line profiles of several species are well matched by the "low" model of \citet{Padovani18}; we adopt that model here as well. We also employ the new description for cosmic ray induced desorption presented in \citet{Sipila21}.

\begin{figure*}
\centering
        \includegraphics[width=1.8\columnwidth]{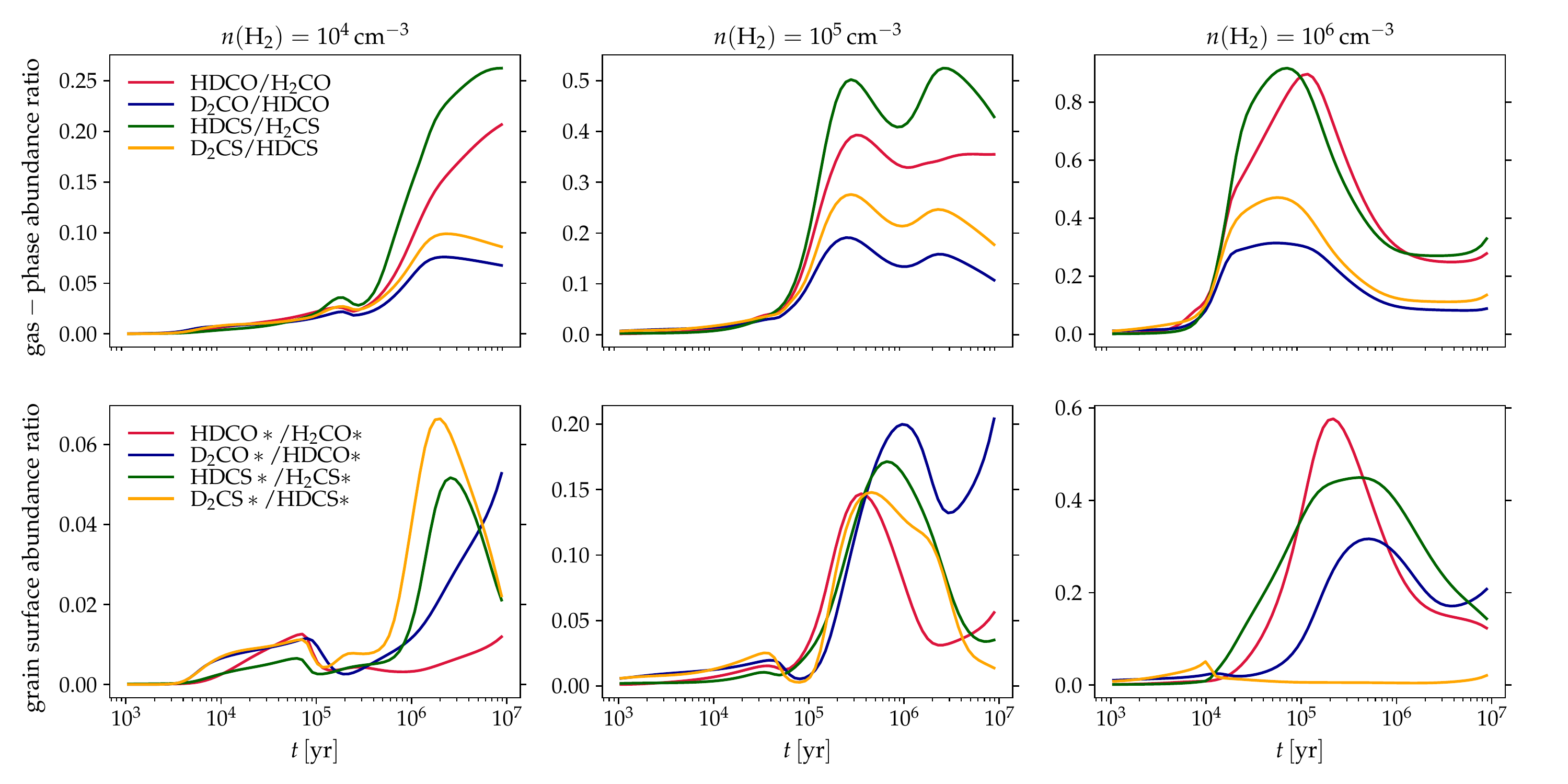}
    \caption{Simulated deuterium fraction ratios of $\rm H_2CO$ and $\rm H_2CS$ as functions of time in the single point model (0D). The top row displays gas-phase abundance ratios, while the grain-surface ratios are displayed on the bottom row. From left to right, the columns correspond to a volume density of $n({\rm H_2}) = 10^4\,\rm cm^{-3}$, $n({\rm H_2}) = 10^5\,\rm cm^{-3}$, or $n({\rm H_2}) = 10^6\,\rm cm^{-3}$, respectively. The asterisks denote grain-surface molecules.}
    \label{fig:dh_ratios}
\end{figure*}

Figure~\ref{fig:dh_ratios} shows the simulated gas-phase and grain-surface deuterium fraction of $\rm H_2CO$ and $\rm H_2CS$ in the single-point chemical models. The values of the various ratios depend on the volume density, but it is evident that in all cases the simulated doubly-to-singly deuterated ratios are clearly below unity, and lie between 0.05 and 0.4. In particular, the ratios are well below unity also on the grain surfaces, which shows that the low gas-phase ratios (as compared to the observations) are not due to inadequate desorption.

\begin{figure}
\centering
        \includegraphics[width=1\columnwidth]{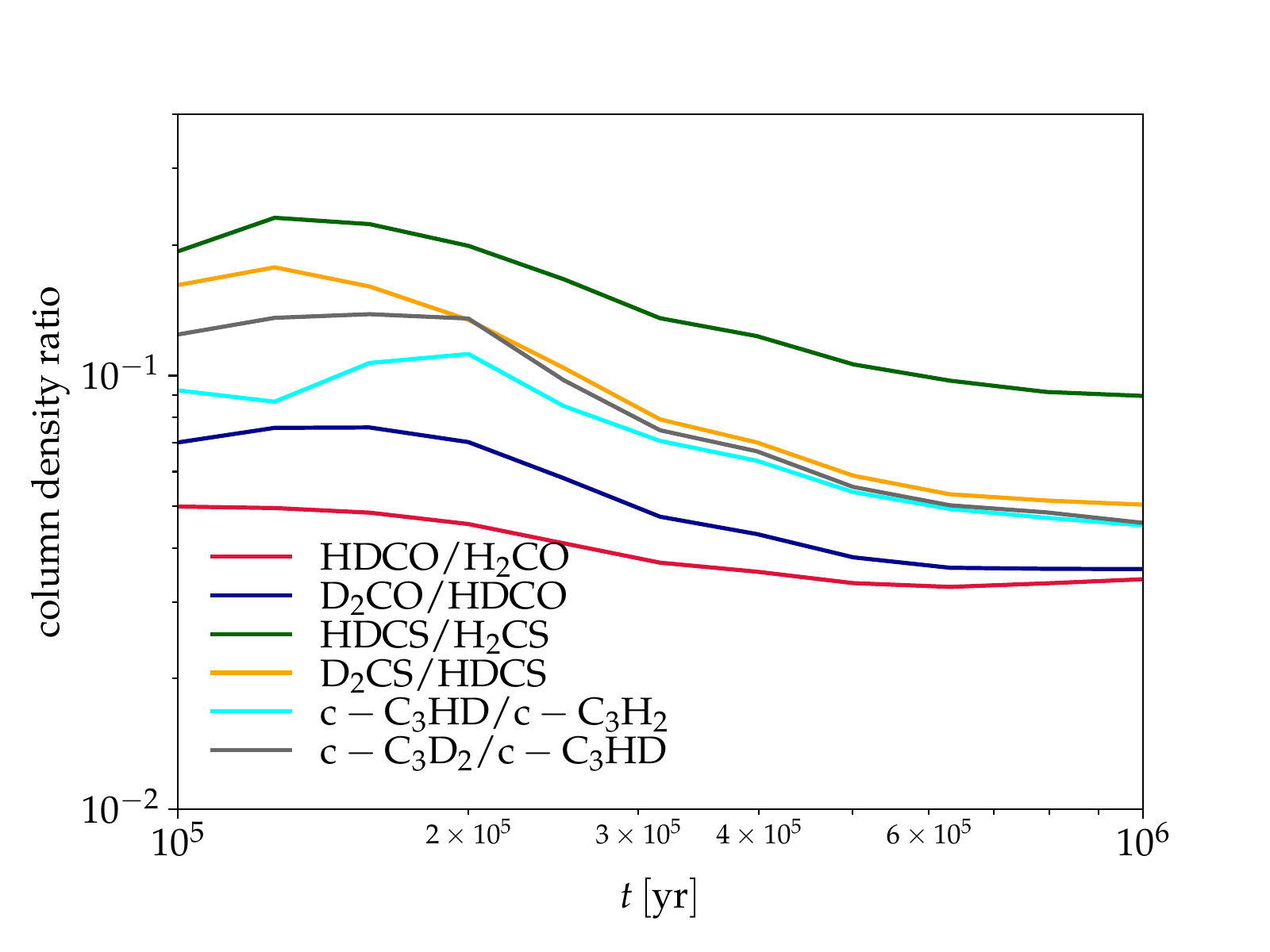}
    \caption{Simulated column density deuterium fractions of $\rm H_2CO$, $\rm H_2CS$, and $c$-C$_3$H$_2$ as functions of time in the L1544 model (1D).}
    \label{fig:L1544}
\end{figure}

The results of the L1544 simulation are shown in Figure~\ref{fig:L1544}, which displays the simulated column density ratios toward the center of the model core, i.e., toward the dust peak in L1544. The column densities have been convolved to the appropriate beam sizes. Comparing the simulated ratios to the observations tabulated in Table~3 shows that the model reproduces well the HDCX/$\rm H_2CX$ ratios (X=O or S), while the amount of double deuteration is again underestimated by the model by an order of magnitude. Note that in the core model, the $\rm D_2CO/HDCO$ ratio is enhanced with respect to the $\rm HDCO/H_2CO$ ratio when compared to the results of the single-point simulations (Fig.\,\ref{fig:dh_ratios}). In the core model, the temperature ranges from $\sim$6 K in the centre to $\sim$20 K in the outer core. The efficiency of deuterium chemistry is sensitive to the temperature, and hence the spatial variations in temperature affect deuteration across the core. The ratios shown in Figure~\ref{fig:L1544} are a result of a line-of-sight integration of the column density which include these spatial variations. For this reason, the results of the core model cannot be compared one-to-one to the single point models, which adopt a constant temperature of 10\,K. The deuterium fractions $c$-C$_3$H$_2$ are slightly underestimated by the model, but lie within a factor of about two of the observed values. In this case too the model predicts an enhancement of the doubly-to-singly deuteration ratio over the singly-deuterated-to-normal ratio\footnote{And the opposite in the single-point models (not shown in Fig.\,\ref{fig:dh_ratios})}.

The simulation results are naturally sensitive to the adopted model parameters. We have tried out several variations of our models in an effort to boost the doubly-to-singly deuterated ratios: 1) gas-phase chemistry only; 2) multilayer ice chemistry; 3) modifications to the branching ratios of surface reactions to promote the formation of doubly-deuterated molecules; 4) decreased activation energies for the formation of D-bearing molecules on the grain surface; 5) switching from complete scrambling to proton hop as the main deuteration mechanism \citep{Sipila19b}. Some of these schemes are able to boost the doubly-to-singly deuterated ratios of $\rm H_2CO$ and $\rm H_2CS$ to a level of $\sim$0.5, but always at the associated cost of increasing the HDCX/$\rm H_2CX$ ratios as well. It remains unknown why the doubly-to-singly deuterated ratios in L1544 are boosted for $\rm H_2CO$ and $\rm H_2CS$, but not for $c$-C$_3$H$_2$, see Table~2. It is however important to note that the main difference between $c$-C$_3$H$_2$ and H$_2$CO or H$_2$CS is that towards cold cores $c$-C$_3$H$_2$ is formed and deuterated only by gas-phase reactions \citep{spezzano13}, while H$_2$CO and H$_2$CS need a combination of reaction in the gas phase and on the surface of dust grains.

\section{Discussion}
\subsection{Deuteration maps}
The N(HDCS)/N(H$_2$CS) deuteration map shown in the left panel of Figure~\ref{fig:H2CS_ratio} presents a peak towards the north-east of L1544. The column density ratio error map is shown in Figure~\ref{fig:error}. 
H$_2$CS is the first molecule whose deuteration peak is not coincident with the dust peak, or in its vicinity within 5000 AU. The deuteration peak of H$_2$CS towards the north-east seems to suggest that its deuteration is more efficient in the outer layers of the core. 
The normal and deuterated isotopologues of H$_2$CS do not necessarily trace the same regions within pre-stellar cores, and we expect H$_2$CS to be present also in the external layers of L1544, while HDCS and D$_2$CS will be efficiently produced only in the inner 6000 AU of the core, the so-called deuteration zone \citep{caselli12}. When moving from the dust peak towards the N(HDCS)/N(H$_2$CS) peak in the north-east, the column density of H$_2$CS shows a steeper decrease compared to the HDCS column density. 
While the H$_2$CS column density drops by 50\%, the HDCS column density drops by only 20\% suggesting that the deuteration peak towards the north-east of L1544 might be a consequence of the steeper drop of the H$_2$CS in the outer layers of L1544, and not the result of a local enhancement in the deuteration of the molecule.
The chemistry in the outer layers of L1544, where we expect only H$_2$CS and not its deuterated isotopologues to be present, is more affected by the uneven illumination onto the core than the inner layers, and in turn also the distribution of H$_2$CS is not expected to be even, but to peak towards the south, the $c$-C$_3$H$_2$ peak, and decrease towards the north-east, the CH$_3$OH peak \citep{spezzano16, spezzano17}. Being the column density a measure of the column of molecules along the line-of-sight, it takes into account the molecules present in the outer as well as in the inner layers of the core, and the column density ratios that we measure at different offsets are affected by the inhomogeneous distribution of the molecule in the different layers of the core. The illumination does not have an effect on the D/H ratio, it only dilutes the N(HDCS)/N(H$_2$CS) ratio towards the South of L1544, where H$_2$CS is more efficiently formed in the outer layers of the core. This is due to the fact that more efficient formation of H$_2$CS is expected in regions where C atoms are not mainly locked in CO (as in the southern region of L1544, rich in carbon-chain molecules; \citealt{spezzano17}).
The same behaviour as H$_2$CS is seen also in CH$_3$OH and H$_2$CO \citep{chacon19}, although H$_2$CS shows the farthest distance of the deuteration peak from the dust peak ($\sim$10000 AU). The deuteration peak for methanol and formaldehyde is shifted by few thousand AU towards the south-west and north-west of the dust peak, respectively, in agreement with the direction of steepest decrease of the column density of the main isotopologue. 
The situation is slightly different for N$_2$H$^+$ and HCO$^+$, whose deuteration maps have been studied in \cite{redaelli19}. 
N$_2$H$^+$ forms directly from molecular Nitrogen, N$_2$, a late-type molecule that is not very abundant in the outer layers of L1544, and as a consequence N$_2$H$^+$ is more abundant towards the center of the core \citep{caselli99, hily-blant10}. N$_2$H$^+$ shows signs of depletion only at the very center of starless cores \citep{bergin02, caselli02, redaelli19}, where N$_2$ also starts to freeze-out. 
Both N$_2$H$^+$ ad N$_2$D$^+$ are centrally concentrated in starless cores, and consequently also the N(N$_2$D$^+$)/N(N$_2$H$^+$) map peaks at the centre of L1544. HCO$^+$ and DCO$^+$ instead are also quite abundant in the cloud surrounding the core, and their column density maps and deuteration map will reflect the overall gas distribution. In the case of L1544, N(HCO$^+$), N(DCO$^+$) as well as their ratio map peaks towards the north-west of the dust peak (see the bottom panel of Figure 8 and Figure 10 in \citealt{redaelli19}).

It is important to note that the N(HDCS)/N(H$_2$CS) ratio is 0.12$\pm$0.02 at the dust peak and 0.27$\pm$0.07 at the deuteration peak towards the north-east, so the difference in the deuteration at the two peaks is only at 2$\sigma$ level. Furthermore, the position of the N(HDCS)/N(H$_2$CS) peak is at the border of the 3$\sigma$ contour of the HDCS column density map. Maps with higher sensitivity and angular resolution are needed in order to make quantitative conclusions on the possible local increase of deuteration fraction of H$_2$CS towards the north-east.

\subsection{Deuteration towards the dust peak}
With the D$_2$CS map not being much extended across the core, it is not possible to draw conclusions about the N(D$_2$CS)/N(H$_2$CS) and N(D$_2$CS)/N(HDCS) deuteration maps.
However, we can compare the single and double deuteration of H$_2$CS with cyclopropenylidene, $c$-C$_3$H$_2$, and H$_2$CO, previously observed in their singly and doubly deuterated isotopologues towards the dust peak of L1544 \citep{spezzano13,chacon19}. The column densities and deuteration ratios are summarised in Table \ref{table:ratios}.
Three main conclusions can be drawn from the numbers reported in Table \ref{table:ratios}: i) in the case of $c$-C$_3$H$_2$, N($c$-C$_3$HD) / N($c$-C$_3$H$_2$) $\sim$ N($c$-C$_3$D$_2$) / N($c$-C$_3$HD) $\sim$ 10$\%$; ii) H$_2$CS is more efficiently deuterated than H$_2$CO; iii) the column densities of the singly and doubly deuterated isotopologues of H$_2$CS and H$_2$CO are the same within errorbars, leading to a D$_2$CX/HDCX $\sim 100\%$ (with X= S or O).
As discussed in Section~\ref{chemical_models}, the deuteration of $c$-C$_3$H$_2$ is reproduced fairly well (factor of $\sim$2) by the chemical models with gas-phase reactions. $c$-C$_3$HD is mainly formed by the reaction of $c$-C$_3$H$_2$ with H$_2$D$^+$ or other deuterated isotopologues of H$_3^+$, followed by dissociative recombination with electrons. $c$-C$_3$D$_2$ is formed in the same fashion from the reaction of $c$-C$_3$HD with H$_2$D$^+$. The formation of H$_2$CO and H$_2$CS instead occur both in the gas phase and on the surface of dust grains (by hydrogenation of CO and CS). 
Despite the similar formation pathways, H$_2$CS is more efficiently deuterated than H$_2$CO as a consequence of the longer time spent on the surface because of its higher binding energy.
Another noticeable difference among the deuteration of these three molecules is that the column densities of the singly and doubly deuterated H$_2$CO and H$_2$CS are, within errorbars, the same, while the column density of the doubly deuterated $c$-C$_3$H$_2$ is 10$\%$ of the column density of the singly deuterated.
A larger D$_2$CO/HDCO ratio (12$\%$) with respect to the HDCO/H$_2$CO ratio (6$\%$) has been observed towards the protostellar core IRAS 16293-2422 B with ALMA observations \citep{persson18}. The results of \cite{persson18} can be reproduced with the gas-grain chemical model described in \cite{taquet14}, where the physical and chemical evolution of a collapsing core is followed until the end of the deeply-embedded protostellar stage (Class 0). The time step that best-reproduces the observations of \cite{persson18} is 1$\times$10$^5$ yr, at the beginning of the Class 0 stage, suggesting that the deuterium fractionation observed for H$_2$CO in IRAS 16293-2422 B is mostly inherited from the pre-stellar core phase.

It is interesting to note that towards the protostellar core IRAS 16293-2422 B, as well as towards the pre-stellar core L1544, the HDCS/H$_2$CS ratio is larger than the HDCO/H$_2$CO ratio, suggesting an inheritance of H$_2$CO, H$_2$CS and their deuterated isotopologues from the pre-stellar to protostellar phase.
We were however unable to reproduce the D$_2$CX/HDCX ratios (with X= S or O) observed towards the pre-stellar core L1544, despite using the same reaction schemes used in \cite{taquet14} that include the abstraction and substitution reactions studied in the laboratory by \cite{hidaka09}. The deuteration efficiency is affected when the chemical model is coupled with a hydrodynamical description of core collapse, instead of using a static physical model as is done here, but even in such a case a D$_2$CX/HDCX = 100\% ratio cannot be reached (see Fig. 12 in \citealt{sipila18}). The deuteration on the surface is boosted significantly when using a three-phase model with respect to the two-phase model, but still fails to reproduce the D$_2$CX/HDCX and HDCX/H$_2$CX ratios observed for H$_2$CO and H$_2$CS towards the centre of L1544.
Our chemical models can reproduce fairly well the HDCX/H$_2$CX for both H$_2$CO and H$_2$CS, suggesting that the reaction network for the formation of the doubly deuterated H$_2$CS and H$_2$CO are not complete yet. Additionally, the exothermicity of the formation of D$_2$CX could be larger than for HDCX, leading to a more efficient reactive desorption for the doubly deuterated isotopologues. Different reactive desorption rates for the different isotopologues are not implemented in our chemical models, where all molecules desorb with a constant efficiency from the grains (1\%). However, as pointed out in Section~\ref{chemical_models}, a higher desorption rate for D$_2$CO and D$_2$CS alone will not be able to reproduce the observations.
\cite{marcelino05} observed H$_2$CS, HDCS and D$_2$CS towards Barnard 1 and found N(HDCS)/N(H$_2$CS)$\sim$N(D$_2$CS)/N(HDCS)$\sim$0.3. The authors used steady state gas-phase chemical models and were able to reproduce the observed column density ratios. However, at the time many surface processes were still poorly known, and were not included.
Efficient H-D substitution reactions on the surfaces may play an important role and more laboratory work is needed to quantify their rates.

\section{Conclusions}
We carried out a comprehensive observational study of the deuteration of H$_2$CS towards the pre-stellar core L1544, and compared its deuteration with other molecules observed towards the same source. 
We find that the N(HDCS)/N(H$_2$CS) deuteration map presents a peak towards the north-east of L1544. Given that the column density of the main species towards the north-east of the core drops faster with respect to the deuterated isotopologue, we suggest that the deuteration peak of H$_2$CS towards the north-east could be a consequence of the steeper drop of the H$_2$CS in the outer layers of L1544. However, deeper integrations and observations at higher angular resolution are needed to draw conclusions on the H$_2$CS deuteration peak towards the north-east of L1544.
The present results imply that the large deuteration of H$_2$CO and H$_2$CS observed in protostellar cores as well as in comets could be inherited from the pre-stellar phase, as suggested by previous works. 
We compared the single and double deuteration of $c$-C$_3$H$_2$, H$_2$CS and H$_2$CO and found that while for $c$-C$_3$H$_2$ both deuteration ratios are $\sim$10\%, for H$_2$CS and H$_2$CO the second deuteration is more efficient than the first one, leading to a similar column densities for the singly and doubly deuterated isotopologues.
We used state of the art chemical models to reproduce the observed column density ratios and found that the deuteration of $c$-C$_3$H$_2$ can be very well reproduced both for the single as well as for the double deuteration, but this is not the case for H$_2$CO and H$_2$CS. Our models can reproduce well the column densities of H$_2$CO, H$_2$CS, HDCO and HDCS, but fail to reproduce the observed large D$_2$CO and D$_2$CS column densities, suggesting that the reaction network for the formation of the doubly deuterated H$_2$CS and H$_2$CO is not complete yet. More laboratory work should be dedicated to study H-D substitution reactions on the surface of dust grains. Also a more efficient reactive desorption for the doubly deuterated isotopologues with respect to the singly deuterated and main isotopologues might play a role.

\section*{Acknowledgments}
S.S. thanks the Max Planck Society for the Independent Max Planck Research Group funding.
AF and GE thank the Spanish MICINN for funding support from  PID2019-106235GB-I00.

{}
%\bibliographystyle{frontiersinSCNS_ENG_HUMS} % for Science, Engineering and Humanities and Social Sciences articles, for Humanities and Social Sciences articles please include page numbers in the in-text citations
%\bibliographystyle{frontiersinHLTH&FPHY} % for Health, Physics and Mathematics articles
%\bibliography{test}

%%% Make sure to upload the bib file along with the tex file and PDF
%%% Please see the test.bib file for some examples of references
\newpage
\begin{appendix}
\section{Spectra and column density maps}
\begin{figure*}
\begin{center}
\includegraphics[width=12cm]{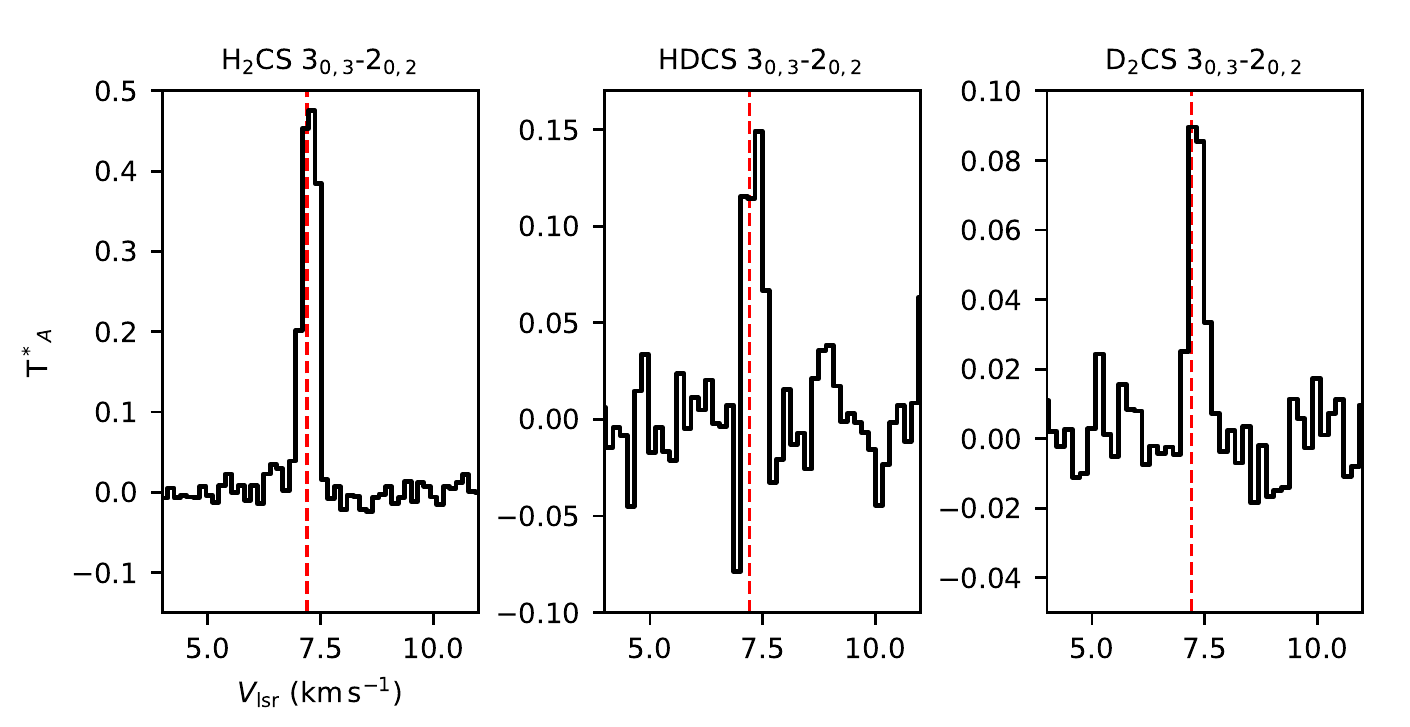}% This is a *.eps file
\end{center}
\caption{Spectra of the 3$_{0,3}$-2$_{0,2}$ transition of H$_2$CS, HDCS, D$_2$CS extracted towards the dust peak of L1544. The vertical dashed red line shows the v$_{LSR}$ of the source, 7.2 km/s.}
\label{fig:spectra}
\end{figure*}

The spectra of the 3$_{0,3}$-2$_{0,2}$ transition of H$_2$CS, HDCS, D$_2$CS extracted towards the dust peak of L1544, i.e. around the centre of the maps shown in Figure~\ref{fig:integrated_intensity}, are shown in Figure~\ref{fig:spectra}.

\begin{figure*}
\begin{center}
\includegraphics[width=18cm]{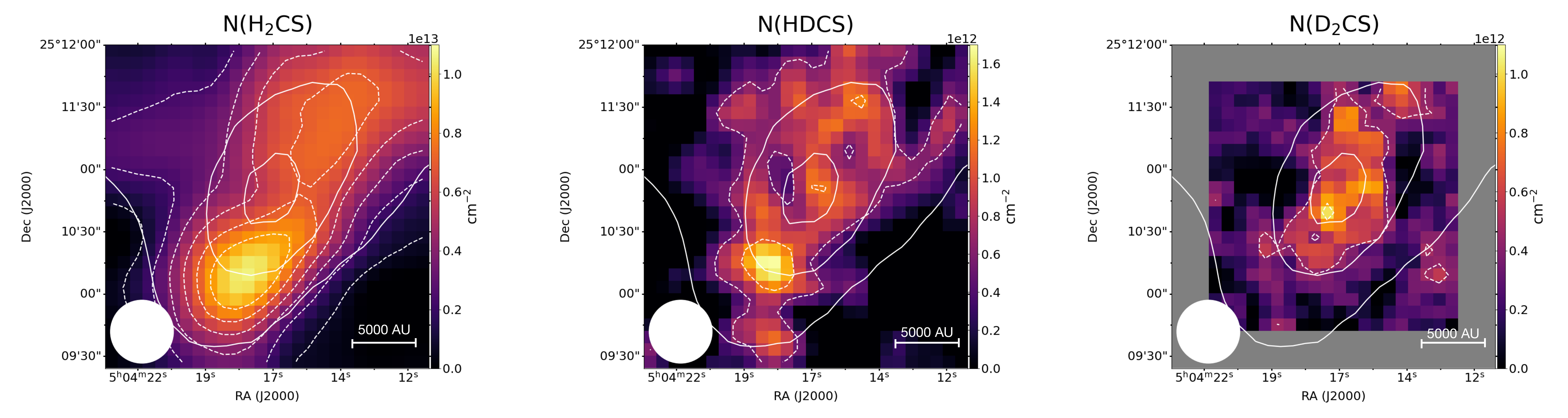}% This is a *.eps file
\end{center}
\caption{Column density maps of H$_2$CS, HDCS, and D$_2$CS towards L1544. The column density has been computed assuming a constant $T_{ex}$ = 12.3, 6.8 and 9.3 K for H$_2$CS, HDCS and D$_2$CS, respectively. The solid white contours are 30\%, 60\% and 90\% of the peak intensity of the N(H$_2$) map of L1544 computed from {\em Herschel}/SPIRE data. The dotted white contours indicate the 3$\sigma$ integrated emission contour with steps of 3$\sigma$ (rms$_{H_2CS}$= 10 mK km s$^{-1}$, rms$_{HDCS}$=12 mK km s$^{-1}$, rms$_{D_2CS}$= 9 mK km s$^{-1}$). }
\label{fig:H2CS_cd}
\end{figure*}

The column density maps of H$_2$CS, HDCS, and D$_2$CS computed assuming $T_{ex}$ = 12.3, 6.8 and 9.3 K for H$_2$CS, HDCS and D$_2$CS, respectively, are shown in Figure~\ref{fig:H2CS_cd}.

\begin{figure*}
\begin{center}
\includegraphics[width=18cm]{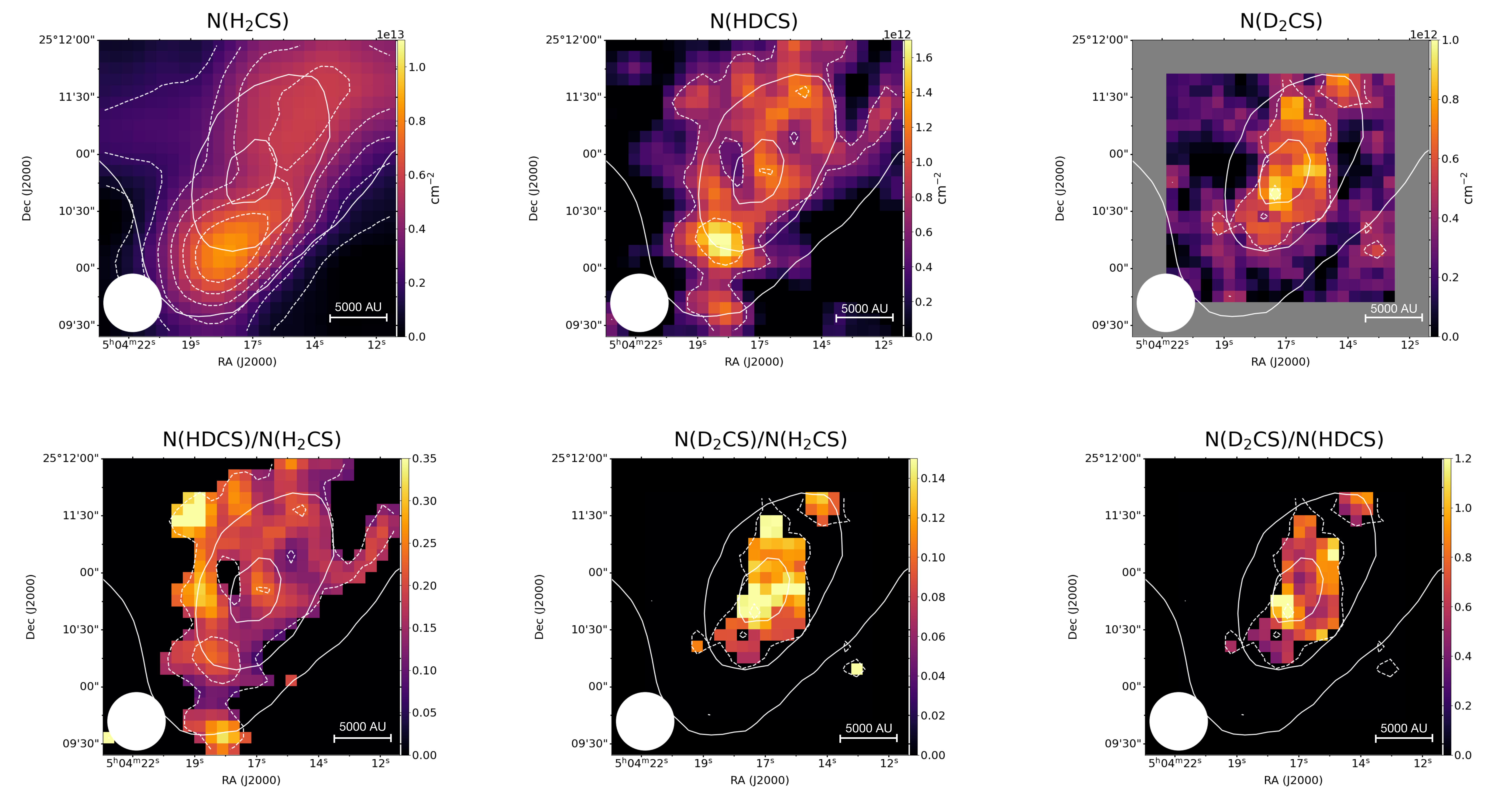}% This is a *.eps file
\end{center}
\caption{Top panels: column density maps of H$_2$CS, HDCS, and D$_2$CS towards L1544. The column density has been computed assuming a constant $T_{ex}$ = 9.3 K for all species. The solid white contours are 30\%, 60\% and 90\% of the peak intensity of the N(H$_2$) map of L1544 computed from {\em Herschel}/SPIRE data. The dotted white contours indicate the 3$\sigma$ integrated emission contour with steps of 3$\sigma$ (rms$_{H_2CS}$= 10 mK km s$^{-1}$, rms$_{HDCS}$=12 mK km s$^{-1}$, rms$_{D_2CS}$= 9 mK km s$^{-1}$).
Bottom panels: deuteration maps of H$_2$CS towards L1544. The column densities have been computed using the same T$_{ex}$ for all species. The column density ratio has been computed only in pixels where both molecules have been observed at least at a 3$\sigma$ level. The solid white contours are 30\%, 60\% and 90\% of the peak intensity of the N(H$_2$) map of L1544 computed from {\em Herschel}/SPIRE data. The dotted white contours indicate the 3$\sigma$ integrated emission contour for HDCS in the left panel, and of D$_2$CS in the central and right panel. 
}
\label{fig:H2CS_cd9.4}
\end{figure*}

The column density maps computed assuming $T_{ex}$ = 9.3 K for all isotopologues, as well as the corresponding deuteration maps, are shown in Figure~\ref{fig:H2CS_cd9.4}. Figure~\ref{fig:H2CS_cd9.4} shows that our main results and conclusions hold even assuming the same T$_{ex}$ for all isotopologues, in fact the N(HDCS)/N(H$_2$CS) peaks towards the north-east of L1544 and the N(D$_2$CS)/N(HDCS) is $\sim$100\%.

\begin{figure*}
\begin{center}
\includegraphics[width=18cm]{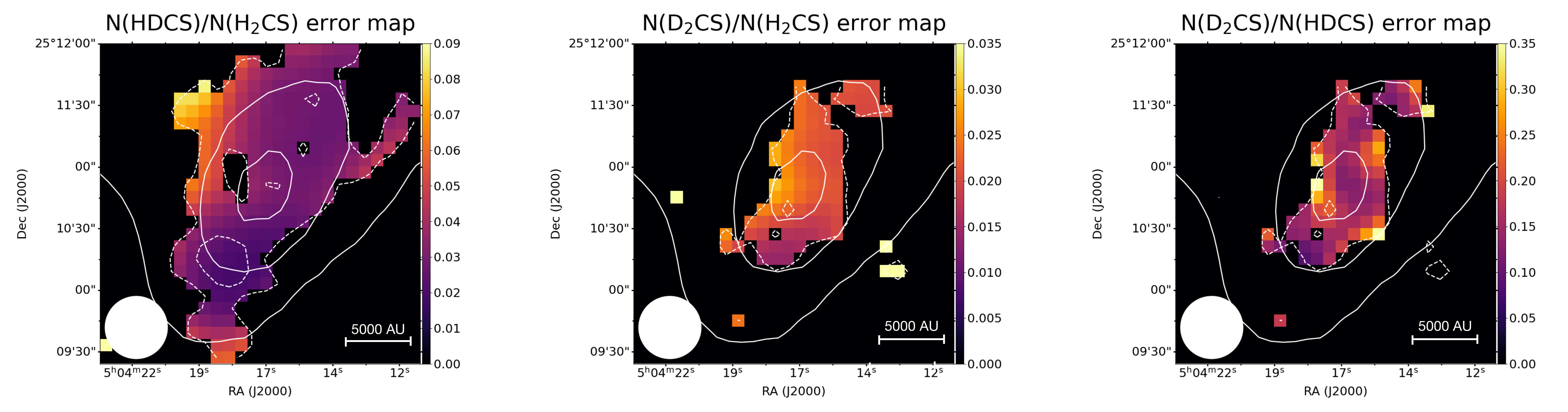}% This is a *.eps file
\end{center}
\caption{Column density ratio error maps. The solid white contours are the 30\%, 60\% and 90\% of the peak intensity of the N(H$_2$) map of L1544 computed from {\em Herschel}/SPIRE data \citep{spezzano16}. The dashed white contours indicate the 3$\sigma$ integrated emission with steps of 3$\sigma$ (rms$_{H_2CS}$= 10 mK km s$^{-1}$, rms$_{HDCS}$=12 mK km s$^{-1}$, rms$_{D_2CS}$= 9 mK km s$^{-1}$).
}
\label{fig:error}
\end{figure*}

The error maps for the column density ratios in Figure~\ref{fig:H2CS_ratio} are shown in Figure~\ref{fig:error}.

\end{appendix}

\end{document}